\documentclass{mn2e}
\usepackage{natbib}
\usepackage{graphicx}
\usepackage{pslatex}

\begin{document}


\title{Erratum: Luminosity function, sizes and FR dichotomy of radio-loud AGN}

\author[C.R. Kaiser \& P.N. Best]{Christian R. Kaiser\thanks{crk@soton.ac.uk} and Philip N. Best\\
}

\maketitle

\begin{keywords} 
errata, addenda -- galaxies: jets -- galaxies: active -- galaxies: luminosity function -- radio continuum: galaxies
\end{keywords} 

\renewcommand{\theequation}{A\arabic{equation}}
\setcounter{equation}{8}

The paper `Luminosity function, sizes and FR dichotomy of radio-loud AGN' was published in Mon. Not. R. Astron. Soc. {\bf 381}, 1548--1560, (2007). Since publication we discovered a mistakes in some of the equations in the appendix which describe the radio source evolution according to \citet{ka96b} and \citet{kda97a}. The mistakes do not alter our conclusions in the original paper in any way. However, to avoid future confusion arising from their application, we correct them here. 

In the expressions for the energy densities of the magnetic field, $u_{\rm B}$, and the relativistic electrons, $u_{\rm e}$, equations (\ref{ue}) and (\ref{ub}), we accidentally swapped the symbols for the energy densities. The correct expressions read
\begin{eqnarray}
u_{\rm e} \left( t_{\rm i} \right) & = & \frac{p \left( t_{\rm i} \right)}{\left( \Gamma _l -1 \right) \left( k + 1 \right) \left( r + 1 \right)}, \label{ue}\\
u_{\rm B} \left( t_{\rm i} \right) & = & \frac{r p \left( t_{\rm i} \right)}{\left( \Gamma _l -1 \right) \left( r + 1 \right)}. \label{ub}
\end{eqnarray}
Subsequently, the expression for the energy density of the magnetic field at the observing time $t$ becomes
\begin{equation}
u_{\rm B} \left( t_{\rm i} \right) = \frac{r p \left( t \right)}{\left( \Gamma _l -1 \right) \left( r + 1 \right)} \left( \frac{t}{t_{\rm i}} \right)^{a_1 \left( 4/3 - \Gamma _l \right) / 4}.
\end{equation}

\setcounter{equation}{14}
The changes also affect the expression for the Lorentz factor of the emitting electrons. We therefore need to replace equation (\ref{fg}) with
\begin{equation}
f_{\gamma} = \sqrt{ \frac{2 \pi m_{\rm e} \nu}{e} \sqrt{\frac{\left( \Gamma _l -1\right) \left( r + 1 \right)}{2 \mu_0 r}}}.
\label{fg}
\end{equation}
\setcounter{equation}{16} Also, the normalisation of the electron energy spectrum changes to 
\begin{eqnarray}
f_n & = & \frac{1}{\left(\Gamma _l -1\right) \left( r + 1 \right) \left( k +1 \right) m_{\rm e} c^2}\nonumber\\
& & \left( \frac{\gamma _{\rm min}^{2-m} - \gamma _{\rm max}^{2-m}}{m-2} - \frac{\gamma _{\rm min}^{1-m} - \gamma _{\rm max}^{1-m}}{m-1} \right)^{-1},
\end{eqnarray}
for $m \ne 2$ and
\begin{eqnarray}
f_n & = & \frac{1}{\left(\Gamma _l -1\right) \left( r + 1 \right) \left( k+1 \right) m_{\rm e} c^2} \nonumber\\
& &\left( \ln \frac{ \gamma _{\rm max}}{\gamma _{\rm min}} - \frac{1}{\gamma_{\rm min}} + \frac{1}{\gamma_{\rm max}} \right)^{-1},
\end{eqnarray}
for $m=2$.\setcounter{equation}{19}

Finally, the factor $f_L$ in the expression for the radio luminosity should be
\begin{equation}
f_L = \frac{2 A^{2 \left( 1-\Gamma _l  \right) / \Gamma_l} r f_{\gamma}^{3-m} f_n}{3 \left( r + 1 \right)} \frac{\sigma_{\rm T} c}{\nu}.
\end{equation}

All other equations in the appendix and the main text remain unchanged. 

\def\newblock{\hskip .11em plus .33em minus .07em}

\bibliography{crk}
\bibliographystyle{mn2e}

\end{document}